\documentclass[a4paper,fleqn,usenatbib]{mnras}

\usepackage{newtxtext,newtxmath}
\usepackage[T1]{fontenc}
\usepackage{ae,aecompl}

\usepackage{graphicx}	% Including figure files
\usepackage{amsmath}	% Advanced maths commands
\usepackage{amssymb}	% Extra maths symbols
\newcommand\tab[1][1cm]{\hspace*{#1}}

\title[Using CNNs to identify gravitational lenses]{Using Convolutional Neural Networks to identify Gravitational Lenses in Astronomical images}

\author[A. Davies]{
Andrew Davies$^{1}$\thanks{E-mail: Andrew.Davies@Open.ac.uk}, Stephen Serjeant$^{1}$, Jane M. Bromley$^{2}$
\\
$^{1}$School of Physical Sciences, The Open University, Walton Hall, Milton Keynes, MK7 6AA, UK\\
$^{2}$School of Computing \& Communications, The Open University, Walton Hall, Milton Keynes, MK7 6AA, UK
}

\date{Accepted XXX. Received YYY; in original form ZZZ}

\pubyear{2018}

\begin{document}
\label{firstpage}
\pagerange{\pageref{firstpage}--\pageref{lastpage}}
\maketitle

\begin{abstract}
The Euclid telescope, due for launch in 2021, will perform an imaging and slitless spectroscopy survey over half the sky, to map baryon wiggles and weak lensing. During the survey Euclid is expected to resolve 100,000 strong gravitational lens systems. This is ideal to find rare lens configurations, provided they can be identified reliably and on a reasonable timescale. For this reason we have developed a Convolutional Neural Network (CNN) that can be used to identify images containing lensing systems. CNNs have already been used for image and digit classification as well as being used in astronomy for star-galaxy classification. Here our CNN is trained and tested on Euclid-like and KiDS-like simulations from the Euclid Strong Lensing Group, successfully classifying 77\% of lenses, with an area under the ROC curve of up to 0.96. Our CNN also attempts to classify the lenses in COSMOS HST F814W-band images. After convolution to the Euclid resolution, we find we can recover most systems that are identifiable by eye. The Python code is available on Github.
\end{abstract}


%%%%%%%%%%%%%%%%%%%%%%%%%%%%%%%%%%%%%%%%%%%%%%%%%%

%%%%%%%%%%%%%%%%% BODY OF PAPER %%%%%%%%%%%%%%%%%%

\section{Introduction}
\label{sec:Introduction}

Gravitational lensing is caused by the mass of a foreground object, such as a galaxy or galaxy cluster, deflecting light from another distant source object, such as a galaxy or quasar. Strong gravitational lensing is rare with only a few systems expected from surveying thousands of objects \citep{Blain1996}. The first strong gravitational lens system, QSO 0957+561, was recognised as such in 1979 when the spectra of two objects were compared and confirmed to be from the same object \citep{Walsh1979}. 
\par
The Jodrell Bank-Very Large Array (VLA) Astrometric Survey (JVAS) \citep{Patnaik1992,Browne1997} and the Cosmic Lens All-Sky Survey (CLASS) \citep{Browne2003,Myers2003} have detected 22 radio loud lensed active galactic nuclei \citep{Chae2003}. Currently the Sloan Lens ACS Survey (SLACS) has provided the most strong lensed systems from a single survey with nearly 100 observed \citep{Bolton2008}. Other sources of strong lenses include the Kilo-Degree Survey (KiDS) \citep{deJong2015} which uses the VLT Survey Telescope at the Paranal Observatory in Chile, the Dark Energy Survey \citep{DarkEnergySurveyCollaboration2005}, and the Subaru Hyper Suprime-Cam Survey \citep{Miyazaki2012}, expected to observed thousands of lenses \citep{Collett2015}. The BOSS Emission-Line Lens Survey (BELLS) have discovered at least 25 strong galaxy$-$galaxy gravitational lens systems
with lens redshifts $0.4 < z < 0.7$, discovered spectroscopically by the presence of higher redshift emission lines within the Baryon Oscillation Spectroscopic Survey (BOSS) of luminous galaxies, and confirmed with high$-$resolution Hubble Space Telescope (HST) images \citep{Brownstein2012}. 
\par
Lensing systems are extremely useful cosmological tools. Lensed systems can be used to constrain the value of the Hubble constant, $H_{0}$, by measuring time delay \citep{Refsdal1964,Kochanek2004}, which occurs because the light from multiple images has taken different paths to reach the observer, introducing a time delay. Cosmological distances are proportional to $c / H_{0}$, meaning $ \Delta t = (1 / H_{0})k$ where $k$ is related to the lens mass model. If a lens model can be found then we can predict $ \Delta t H_{0}$ and infer $H_{0}$. Gravitational lensing is independent of the lensing object's luminosity and depends only on the mass of the lens object and the geometry of the source and the lens relative to the observer. This makes lensing a unique tool for analysing mass distribution in the foreground lens \citep{Treu2002}. Using this dependence on mass alone, and combining mass models from mass$-$luminosity analysis, the baryonic and dark matter mass of the galaxy can be mapped to find dark matter substructure \citep{Vegetti2012,Metcalf2001}. Gravitational lensing conserves surface brightness (a consequence of Liouville's Theorem) but not the angular size of the source object \citep{Marchetti2017}, causing a magnification of the source object's flux, if the image of the source is enlarged. This enables the observation of fainter galaxies which would otherwise be missed, including galaxies at high redshifts \citep{Claeskens2006,Jackson2011,Wyithe2011,Marchetti2017}.
\par
Future telescopes are expected to observe many more strongly lensed systems. The Euclid telescope \citep{Laureijs2011} and the Large Synoptic Survey Telescope \citep{LSSTScienceCollaboration2009}, will bring the total number of systems above $10^{5}$ \citep{Oguri2010,Collett2015}. Euclid will map three-quarters of the extragalactic sky with 0.2 arcsecond resolution to 24 AB magnitude \citep{Amendola2018}. Another project, the Square Kilometer Array (SKA) \citep{Rawlings2011}, will take observations at a  resolution of 2 milliarcseconds at 10 GHz, and 20 milliarcseconds at 1 GHz \citep{Perley2009}. The lensing surveys using SKA are expected to observe $\approx 10^{5}$ new radio$-$loud gravitational lenses \citep{McKean2015,Serjeant2014}. 
\par
There is currently a shift in the methodology for detecting strongly lensed systems in astronomical images as numbers of lens candidates becomes much larger. Traditionally most images were found by eye. 112 lens candidates, and at least 2 certain lenses, were found in a HST legacy programme, looking at the COSMOS field \citep{Faure2008,Jackson2008}. But not all searching by eye has been carried out by people working in strong gravitational lensing. The public have been tasked with finding new lens candidates through the Space Warps citizen science project \citep{Marshall2016}. Space Warps made use of volunteers analysing 430,000 images by eye to look for lensing features, via an online webpage using the Zooniverse platform. Tens of new lens candidates have been identified with the help of these volunteers, using large ground$-$based surveys, e.g. the Canada$-$France$-$Hawaii Telescope Legacy Survey \citep{More2016}. But with the growth of survey size there will be too many candidates to be examined by eye. 
\par
There have been several successful methods of computational searches for lenses. Arcfinder \citep{Seidel2007}, uses pixel-grouping methods to attempt to find cluster$-$scale lens systems. Ringfinder \citep{Gavazzi2014}, searches for blue residuals surrounding early$-$type galaxies using multi$-$band data, also there are several programs to find arc$-$like shapes \citep{Lenzen2004,More2012}. 
\par
In recent years there has been a rise in machine$-$learning methods to detects lenses. 56 lens candidates were found in the KiDS dataset using a convolutional neural network \citep{Petrillo2017}. Machine learning methods rely on large datasets in order to train and learn, something which has become available in recent years. Once a machine has been trained, thousands of images can be classified in seconds. Speed is an important factor due to the expected $10^{9}$ images from Euclid \citep{Collett2015}. The use of citizen science could be used to create a dataset of images for machine learning techniques to train on, once trained, citizen science can then be applied again to examine the output images from the machine learning technique.
\par 
In section 2 we discuss neural networks and why we use them for this problem. In section 3 we discuss the simulated data we have used. In section 4 we discuss our convolutional neural network and how we trained it, and the results are discussed in section 5. The Python code is available at Github \footnote{https://github.com/A-Davies/LensCNN}.

\section{Why use Neural Networks}
\label{sec:Why use Neural Networks}

Computers are very effective at tasks with a limited set of rules, such as chess. However, humans are still often better at real world tasks which cannot easily be described by a set of rules, e.g. recognising objects. Artificial Neural Networks (NNs) are loosely inspired by how the brain works. They are made from simple computing elements with multiple inputs and one output analogous to a brain made up from neurons with dendrites and cell body receiving the inputs and axon outputting a signal. Like the brain, a NN can modify strengths of connections learnt from examples. Humans have evolved to be very fast and accurate at recognising objects, order of 100 ms \citep{Thorpe1996}. NNs, particularly Convolutional Neural Networks (CNNs) are the best available techniques in some tasks, e.g. translation, visual object recognition \citep{LeCun2015}. With current technology, a trained CNN can also perform these tasks faster than humans. 

\begin{figure}
	\includegraphics[width=\columnwidth]{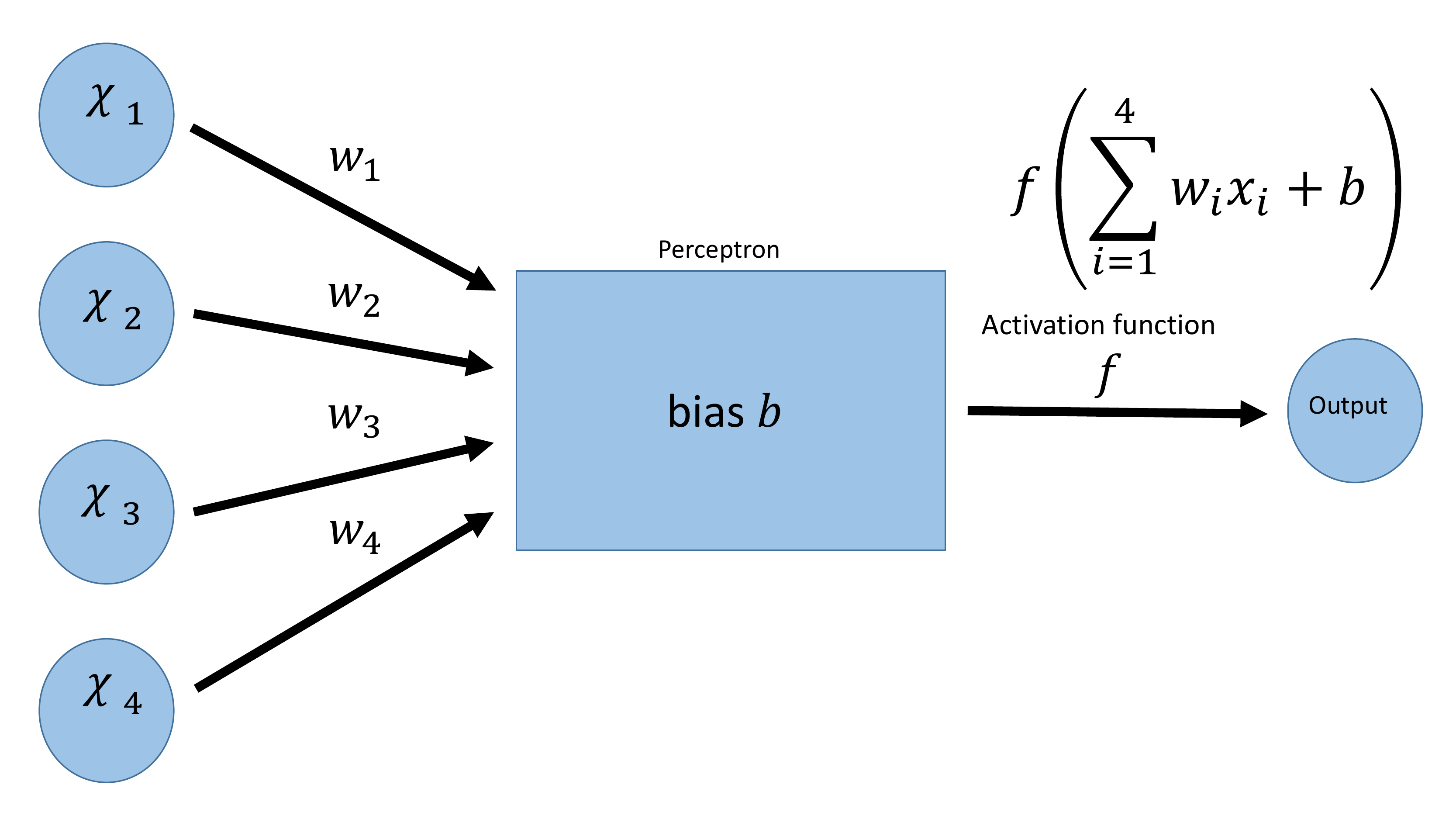}
    \caption{A perceptron with inputs $x_{1}, x_{2}, x_{3}, x_{4}$, weights $w_{1}, w_{2}, w_{3}, w_{4}$, bias $b$, and output calculated from the activation function $f$, together with the product of the weights and input and bias added.}
    \label{fig:Perceptron}
\end{figure}

\begin{figure}
	\includegraphics[width=\columnwidth]{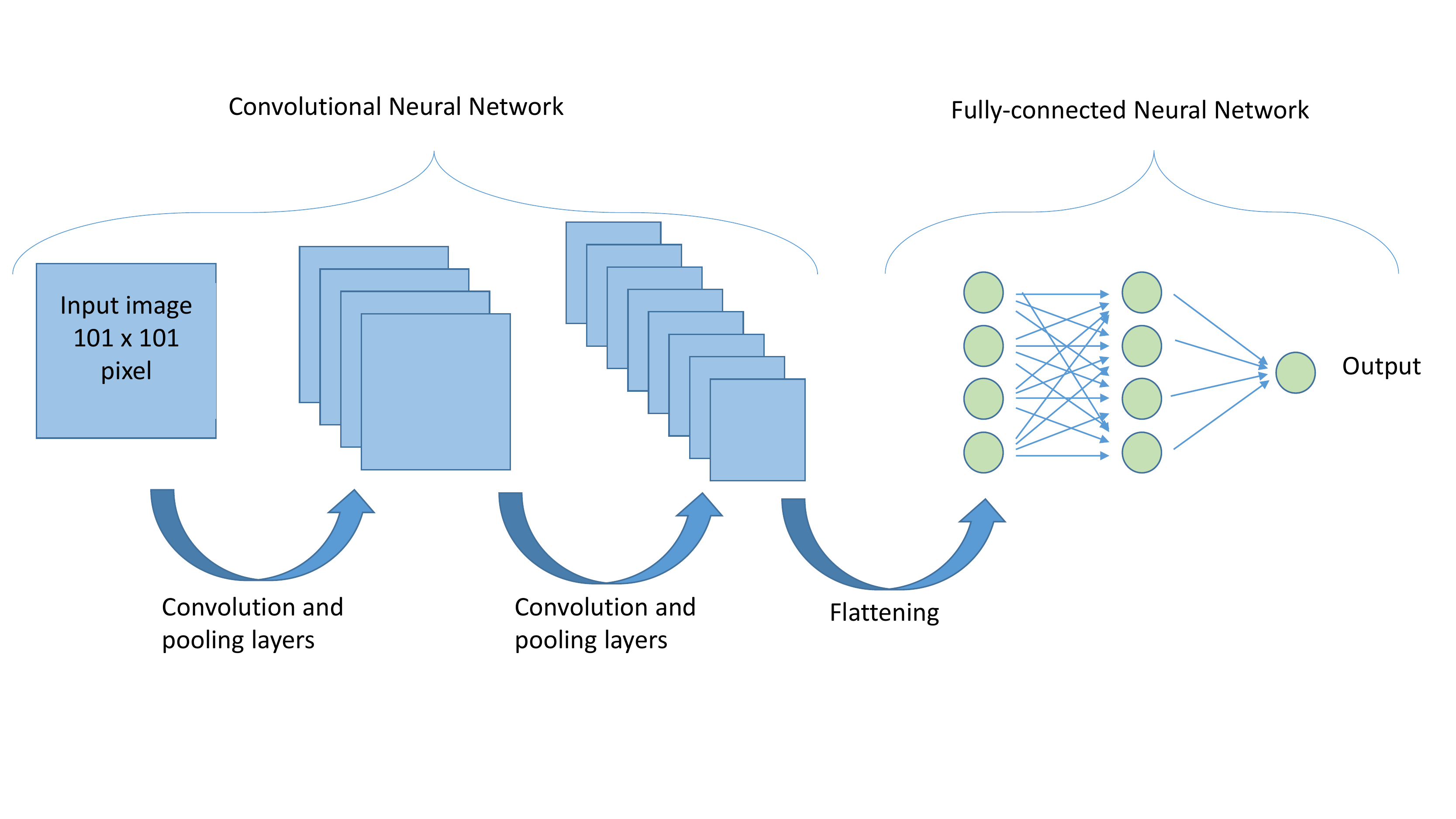}
    \caption{Diagram showing the architecture of the network and where the convolution layers are in respect to the fully-connected neural network. Not all layers shown.}
    \label{fig:Network}
\end{figure}

Neural Networks are built from individual artificial neurons called perceptrons. A perceptron is designed to simulate the role of a biological neuron, but with the advantage that a mathematical function can be used for activation of the neuron \citep{Aggarwal2014}. A perceptron takes a number of inputs, applies a weight to each, sums these products, applies a bias, and then is used as input to an activation function, which then gives the perceptron's output. Perceptrons can accept multiple inputs and apply different weights to each individual input. A perceptron with example inputs and outputs can be seen in  Fig.~\ref{fig:Perceptron}. Individual scalar inputs are grouped together as a 1-d vector. If we let $\textbf{x}$ be the 1-d vector of inputs, $\textbf{w}$ be the 1-d vector of weights associated with this perceptron, $b$ be the bias, and $f$ be the activation function, then the output $z$ is calculated using: 
\begin{equation}
z = f(\sum_{i=1}^{n} w_{i} x_{i}  + b)
\label{eq:Perceptron}
\end{equation} 
where $n$ is the number of inputs to the perceptron. In a NN, many perceptrons are grouped together to form a layer containing a few perceptrons to thousands. Layers use the outputs from previous layers as their inputs. 
\par
Neural Networks have to be trained before they can be used. For classification problems this involves passing many images of a known classification through the network in a process called supervised learning. The network classifies each image, and this output is combined with the true classification in a function called the loss function. A high value for the loss function means many images were incorrectly labelled by the network. The goal of training the network is to minimise the loss function over a number of passes of the training data. Each pass of the data is known as an epoch. After each epoch, the weights and biases are changed using Stochastic Gradient Descent (SGD), in order to minimise the loss function thereby increasing the rate of correct classifications from the network. The rate at which SGD changes the weights and biases is controlled by a variable called the learning rate. The is a linear parameter which controls by how much the weights and biases are changed. Using Adam \citep{Kingma2014} instead of only SGD allows the learning rate to change as the network learns. Initially the learning rate starts off high, and decreases as the network becomes more accurate and the loss function value decreases. A small subset of images are used for data validation. After every epoch the validation images are classified and validation loss is recorded, calculated the same way as the training loss. The network is not trained on the validation data, so no changes are made to the weights and bias. Validation is done to prevent over-training the network. Training is stopped once the validation loss has reached a minimum. The validation loss will increase after this point as the network becomes over-trained, and this extra training is detrimental to classifying new datasets. After training has been completed, new images can be classified by passing the image through the network and obtaining a classification. Classifying an image makes no changes to the network parameters. Batch training means that the weights and biases are updated after seeing only a fraction of the training set. The batch number is typically small compared to the training number. Batch training is used to speed up training since the weights and biases are changed after each batch instead of at the end of each epoch.
\par
CNNs are a subset of NNs which use convolutional layers in the network for feature recognition or classification. An example architecture of a CNN can be seen in Fig.~\ref{fig:Network}. A convolutional layer involves a kernel being convolved with an input image in order to make a feature map. Often the convolution layer has several different kernels for the same image, meaning several different feature maps are given as output. The image kernel can either be pre-determined, or can be another parameter that the network trains and optimises. Different layers may have different image kernel sizes, as well as different sized image outputs. Between convolution layers, pooling layers are often inserted. A pooling layer is designed to greatly reduce the number of pixels in an input image to speed up training and reduce the number of parameters. Pooling is generally done one of two ways, max pooling or mean pooling. Both methods look at a small section of the image, say a $2 \times 2$ section, and reduces this to one pixel value by either finding the maximum value in the $2 \times 2$ square or by finding the mean. The output is then a reduced image with fewer pixels than the input. Pooling is done to reduce the number of variables whilst trying to keep as much spatial information as needed \citep{Mallat2016}. The convolution layers in a CNN are designed to process visual information hierarchically, with earlier layers finding more basic features, and later layers building on what layers before have found to create more complicated features within the image. This is how a network can go from seeing individual pixel values to finding complicated features, such as a face. After the convolutional layers, the network will have a layer to flatten the output from the final layer into a 1-d vector to be used as input for a layer composed of perceptrons. A layer in a network made from perceptrons that each use every output from the previous layer as input is known as a fully-connected or dense layer. All layers in NN apart from the input and output layer are known as hidden layers.
\par
CNNs have been used to solve classification problems such as digit recognition with the MNIST database, a collection of 70,000 $32 \times 32$ grey scale images of handwritten digits. The problem is to classify these as the digits 0 through 9. Using CNNs gives a solution with an error rate of 0.23\% \citep{Ciresan2013}. CNNs have also been successful in object recognition in images. The CIFAR-10 database consists of 60,000 $32 \times 32$ colour images in 10 classes, such as truck, bird and dog, with 6000 images per class. The error rate in this classification problem when using CNNs is lower than 4\% \footnote{http://rodrigob.github.io/are\_we\_there\_yet/build/classification\_datasets\_results.html}. An astronomy classification problem where CNNs have been used is classifying images as a star or galaxy. Here the best network had an error rate of only 0.29\% for galaxies \citep{Kim2007,Dieleman2015}. 

\section{Methods}
\begin{figure}
	\includegraphics[width=\columnwidth]{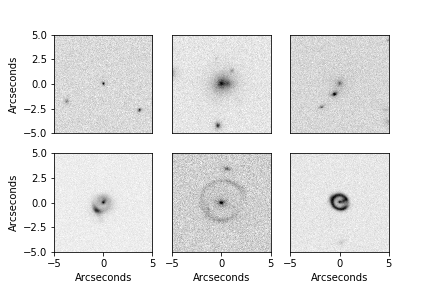}
    \caption{Samples from the 100,000 Euclid VIS instrument simulated images. The top row of images do not contain lenses, while the bottom row contains lenses.}
    \label{fig:EuclidExamples6}
\end{figure}

\begin{figure}
	\includegraphics[width=\columnwidth]{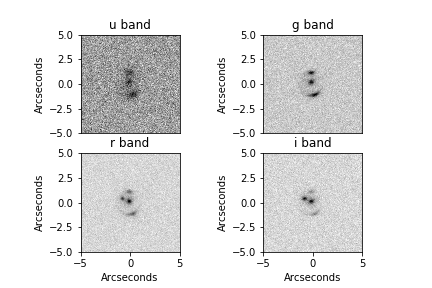}
    \caption{A sample from the 100,000 KiDS-like simulated images. The images are labelled above by their corresponding wavelength band. This example does contain lensing.}
    \label{fig:KiDSExamples4}
\end{figure}

\medskip
The task of finding strong gravitational lenses in large datasets is a problem within Euclid. Feature recognition by eye will not be fast enough to cope with the amount of data received. The Strong Lensing group within the Euclid consortium set up the Euclid Strong Lensing challenge. This was a challenge aimed at developing machine learning techniques to classify images as to whether containing a lens or not. The simulated data was provided by the Bologna Lens Factory \footnote{https://bolognalensfactory.wordpress.com/}. The images are producing using GLAMER code \citep{Metcalf2014}, which uses galaxies from the Millennium Simulation \footnote{https://wwwmpa.mpa-garching.mpg.de/galform/virgo/millennium/} and real galaxies from KiDS (Kilo Degree Survey) as foreground lenses and background sources to produce the simulated images. More details of the lensing process can be found in \citep{LensChallenge}. The simulated images are provided as $101 \times 101$ pixel sized images, centred on the foreground galaxy, either containing a lensed source or not. \textbf{In total 200,000 images were provided, 100,000 were Euclid VIS-like images, with $\approx$ 40,000 lenses and 100,000 were KiDS-like images, with $\approx$ 50,000 lenses}. The Euclid-like space images are single band images, very broad band (r+i+z), whereas the KiDS-like images have four bands; $u, g, r, i$. The images have an image resolution of 0.2 arcsecond, meaning each is a $10 \times 10$ arcsecond square image. Examples of the Euclid VIS-like images can be seen in Fig.~\ref{fig:EuclidExamples6}, and an example of the 4 KiDS-like bands can be seen in Fig.~\ref{fig:KiDSExamples4}. However our close scrutiny of the simulated images uncovered some unphysical examples. The COSMOS lenses \footnote{http://wwwstaff.ari.uni-heidelberg.de/mitarbeiter/cfaure/cosmos/}, were used as a comparison to test the simulations against. Examples of the COSMOS lenses can be seen in the Appendix. The band used for the COSMOS images is the HST F814W wide band. F814W covers the longer wavelength half of the VIS throughput. Visual inspection of the VIS and smoothed COSMOS images shows qualitatively similar features. Therefore, we argue that the COSMOS data set is an appropriate and interesting test of our VIS-trained network.

\begin{figure}
	\includegraphics[width=\columnwidth]{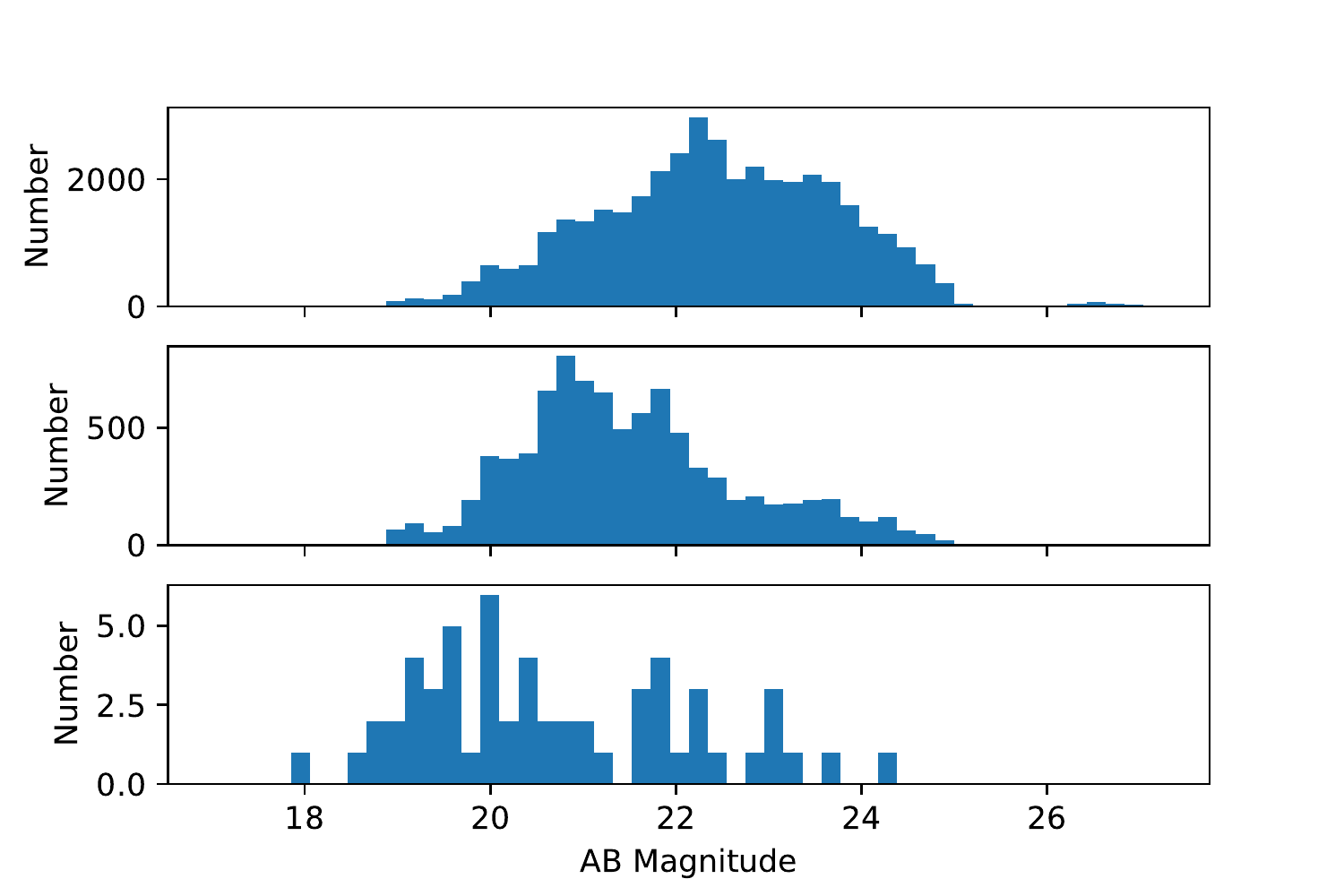}
    \caption{Histograms showing the AB magnitude across 3 datasets. Top: Original 100,000 Euclid VIS-like images, Middle: Subset of Euclid VIS-like images designed to have the same distribution as the COSMOS lenses, Bottom: The COSMOS lenses.}
    \label{fig:COSMOSCut}
\end{figure}

By comparing histograms of the Einstein radius and lens magnitudes of both the simulated Euclid VIS-like images and the real COSMOS lens images, it was found that many of the simulations had unrealistically large Einstein rings, and that the Euclid VIS-like and KiDS-like simulations were fainter than the COSMOS lenses. Because of this, images with Einstein radii greater than 4 arcseconds, and lenses towards the faint end of the Euclid VIS-like and KiDS-like datasets have been removed in order to create a more representative subset. Histograms showing removal of some of the Euclid VIS-like images to make the dataset more COSMOS-like can be seen in Fig.~\ref{fig:COSMOSCut}. A similar histogram also shows that by removing the larger Einstein radii, the Euclid VIS-like images have an Einstein radius distribution similar to that of COSMOS. The same has been done for the KiDS-like dataset. In total we now have 4 datasets for training and 5 for testing which can be seen in Table.~\ref{tab:Datasets}

\begin{table*}
	\centering
	\caption{Table describing the contents of each dataset.}
	\label{tab:Datasets}
	\begin{tabular}{lcr}
		\hline
		Type & Description & Number of lenses\\
		\hline
		Euclid-VIS like simulations & 100,000 single-band $1 \times 101 \times 101$ simulated images & 39975\\
		KiDS-like simulations & 100,000 multi-band $4 \times 101 \times 101$ simulated images & 49862\\
		Euclid-VIS like simulations with COSMOS distribution & 68,923 single-band $1 \times 101 \times 101$ simulated images & 24029\\
		KiDS-like simulations with COSMOS distribution & 60,144 multi-band $4 \times 101 \times 101$ simulated images & 29960\\
		COSMOS lenses & 65 single-band real lenses cropped to $101 \times 101$ images & 65\\
		\hline
	\end{tabular}
\end{table*}

\section{Training to find lenses}
\medskip
Four networks have been built, two designed to work with KiDS-like images with 4 filter bands as input, and two to work with Euclid VIS-like data with a single filter input. They have the same architecture, but have been trained on different data using datasets 1 through to 4 from  Table.~\ref{tab:Datasets}. The networks have been built and trained in Python 2.7 using the neural network library Keras \footnote{https://keras.io/}. \textbf{Keras runs on top of either Theano \footnote{http://deeplearning.net/software/theano/}, TensorFlow \footnote{https://www.tensorflow.org/} or CNTK \footnote{https://github.com/Microsoft/cntk} backends}. We used Theano. The CNN architecture used here has been inspired by the work of \citep{Petrillo2017}. The network architectures can be seen in Table~\ref{tab:NetworkLayers}. 
%HeNormal 
Robust initialisation (HeNormal)
is used to initialise the weights as this speeds up network convergence \citep{HeNormal}. The networks have 4 convolutional layers initially, with $2 \times 2$ max-pooling incorporated twice after the first two convolutional layers. After the convolutional layers, the 2D feature maps that have been made in the final convolutional layer are flattened into a 1D vector to be used as input into the dense layer of fully connected neurons. The final layer is a classification layer, where the network gives each image a classification between 0 and 1. This number can been seen as a probability that the image is a lens. The CNN is trained on a set of 75\% of the labelled images from the dataset, using a process of batch training with a batch size of 500. 5\% of the dataset is used for data validation, to avoid over-training. \textbf{Throughout our networks we used ReLU (Rectified Linear Units \citep{Nair2010}) for the activation functions and binary cross-entropy was used as the loss function}.

\begin{table*}
	\centering
	\caption{Table shows the architecture of the 4 networks, Euclid VIS-like and KiDS-like. What each layers contains, as well as each layers initial weights and biases. The final sigmoid layer gives an output between 0 and 1. \textbf{For the convolutional and max-pooling layers a stride length of 2 was used, meaning each pixel was only used once in pooling, padded with the same edge pixels where required.}}
	\label{tab:NetworkLayers}
	\begin{tabular}{lccccr}
		\hline
		Type of layer & Layer contains & Initial weights & Initial bias\\
		\hline
		Convolutional & 8 ($15 \times 15$) image kernels & HeNormal & Zeroes\\
		Max-pooling & Pooled over each ($2 \times 2$) square & - & -\\
		Convolutional & 8 ($15 \times 15$) image kernels & HeNormal & Zeroes\\
		Max-pooling & Pooled over each ($2 \times 2$) square & - & -\\
		Convolutional & 16 ($5 \times 5$) image kernels & HeNormal & Zeroes\\
		Convolutional & 16 ($5 \times 5$) image kernels & HeNormal & Zeroes\\
		Flatten & Convert image maps into 1-d vector & - & -\\
		Dense & 512 fully-connected neurons & HeNormal & Zeroes\\
		Dense & 1 sigmoid output neuron & HeNormal & Zeroes\\
		\hline
	\end{tabular}
\end{table*}

\section{Results and Discussion}
\begin{figure}
	\includegraphics[width=\columnwidth]{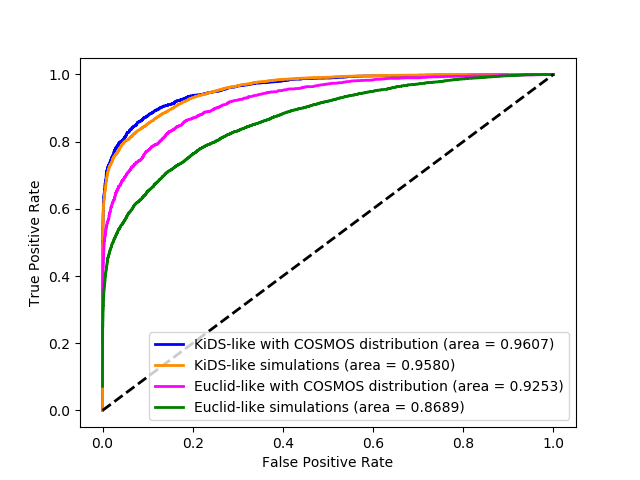}
    \caption{Receiver operating characteristic curves for Euclid network (top) and KiDS network (bottom) with the area under the curve shown. The black dashed line indicates the curve for random choice.}
    \label{fig:JointROC}
\end{figure}

\medskip
An image is judged to contain a lens if the classification from the network is above or equal to $0.5$, conversely if the classification is below $0.5$ the image is judged to not contain a lens. This value can be increased to give a more accurate classification, causing the number of false positives to decrease. However it also means that more lenses are misclassified. 20\% of each dataset is passed through the appropriate network to be classified; this is the test set. The scores for each network can be seen in Table.~\ref{tab:NetworkPercentages}. By looking at the percentage of images classified correctly, and the percentage of lenses and non-lenses classified correctly, the KiDS-like networks are more successful than the Euclid VIS-like networks. This is not surprising since the KiDS-like images have 4 image bands compared to the single band of the Euclid VIS-like images. This would imply that colour information from the multiple bands of the KiDS-like images has been helpful in classifying the lenses correctly, as one would expect as it helps greatly when classifying by eye. A test of this (which we will conduct in future work) would be to compare single-band KIDS-like images with multi-band KIDS-like images. In both the COSMOS-like datasets, the percentage of lenses correctly identified is less than the original datasets. This is probably because the images with the largest rings (> 4 arcseconds Einstein radius) have been removed to make the dataset. These images are easy to identify as lenses and so removing them decreases the success rate for lenses. However non-lens classification success has increased with the same number of non-lens images involved. Although both the datasets have a similar overall success rate, the network trained on the Euclid VIS-like dataset performed significantly worse at recognising the images containing lenses. The difference in overall performance is most clear by looking at the Receiver Operating Characteristic (ROC) curves for the datasets together in Fig.~\ref{fig:JointROC}. A true positive is when a true image, one with a lens, is classified as such, while a false positive is when a false image, one without a lens, is classified as having a lens. \textbf{The area under the ROC curve determines the results, 1 being the score for all classifications correct, 0 the score for all classifications incorrect, and 0.5 being the result of random selection}. True negatives and false negatives are defined similarly. Fig.~\ref{fig:JointROC} confirms that the KiDS-like dataset performed best, although the network only improved slightly using a subset of the images, unlike the Euclid dataset which improved significantly by removing images from the dataset.
\par
As well as using simulated data, we tested on 65 real images from COSMOS. These are single-band images made from larger image cutouts and modified to have the same PSF as the Euclid images by convolving with a suitable kernel. The full width at half maximum (FWHM) of Euclid squared is equal to the FWHM of COSMOS squared plus the FWHM of the kernel squared;
\begin{equation*}
    FWHM(Euclid)^{2} = FWHM(COSMOS)^{2} + FWHM(kernel)^{2}
\end{equation*}
After convolution to match the Euclid PSF, the COSMOS images also had their pixels resampled to match the Euclid pixel scale. Images of the COSMOS lenses before and after applying the convolution can be seen in the Appendix. The resulting images were also $101 \times 101$ pixels. Having only 65 available images meant that training on these images was not a possibility, but testing them with the trained Euclid-like networks was. The results can be seen in Table.~\ref{tab:COSMOSResults}. The scores for both networks were very low, and can be expected after training on a different type of image. Every COSMOS lens that has been classified incorrectly is a false negative. All of the images from the COSMOS survey can be seen in the Appendix. \textbf{Nevertheless, our network recovers 16/31 of the lenses identifiable by eye at the Euclid resolution (see Appendix), and 8/34 of the lenses that cannot be identified by eye}. Although our network at Euclid resolution only recovers $20\%$ of the lenses known to exist at HST resolution, this is in itself quite interesting: it implies that the detectability of lensing is a very strong function of angular resolution. Roughly doubling the angular resolution (from Euclid to HST resolution) results in roughly a five-fold increase in the numbers of detectable lensing systems. 
%The results from the COSMOS lenses, only scoring 20\% implies there exists $\approx5 \times$ more lenses just below Euclid resolution limit. 
%By looking at the COSMOS lenses in the Appendix, one can see that by eye, the images where a Euclid-like kernel has been applied are far more difficult for identifying lenses.
\par
Even though the KiDS-like subset showed a great deal of success, there are still things to be wary of: all the images used in this work in training and testing are simulated images, but real images may not be classified as accurately, which can be seen by looking at the results from the COSMOS images. The percentage of images containing a lens is much higher in these simulated cases than for real data. 
\par
Once implemented with real data, the number of lenses observed will increase. This in turn will increase the number of rare lens systems found, such as double-source plane lens systems \citep{CollettandAuger2014}. These rare systems can be used to constrain the dark-energy equation of state parameter $w$ \citep{Gavazzi2008}. Lens models can be made from the systems observed, and when coupled with visible images can infer dark matter substructure within the lensing galaxy. 
\par
Future work will include training and testing the networks on updated simulations incorporating cluster lenses and Eulcid's grism data. The problems from these first simulations have been noted so that the next training set will not include large Einstein rings (> 4 arcsecond) and will add more complex images to classify, such as face on spiral galaxies. A further complication to be modelled is the diversity of non-lensed interloper populations, such as polar ring galaxies, or galaxies with tidal tails. The most effective approach here may be to degrade HST images to an appropriate Euclid-like resolution. However, the network presented here will be an excellent starting point for training quickly on such a more comprehensive training set, because CNNs can be efficiently adapted to new but similar application domains \citep{Relearn}. The networks will also be trained and tested on different simulations from \cite{Collett2015}.

\begin{table*}
	\centering
	\caption{This table shows the percentage of images classified correctly, for lenses, non-lenses and total correct.}
	\label{tab:NetworkPercentages}
	\begin{tabular}{lccr}
		\hline
		Images & Lenses Correct (\%) & Non-Lenses Correct (\%) & Total Correct (\%)\\
		\hline
		Euclid VIS-like simulations & 60.32 & 93.26 & 80.13\\
		KiDS-like simulations & 88.17 & 86.82 & 87.49\\
		Euclid VIS-like simulations with COSMOS distribution & 56.14 & 98.86 & 93.33\\
		KiDS-like simulations with COSMOS distribution & 76.83 & 97.46 & 93.62\\
		\hline
	\end{tabular}
\end{table*}

\begin{table*}
	\centering
	\caption{Table containing the classification results of the COSMOS lenses on two differently trained networks. Note all 65 images were lenses.}
	\label{tab:COSMOSResults}
	\begin{tabular}{lr}
		\hline
		Trained Dataset & Lenses Correct\\
		\hline
		Euclid VIS-like simulations & 18 (27.69\%)\\
		Euclid VIS-like simulations with COSMOS distribution & 15 (20.00\%)\\
		\hline
	\end{tabular}
\end{table*}

\section{Conclusions}
\medskip
A well designed CNN can be used with future observations from Euclid and other similar surveys as they are demonstrably successful on simulated data. Making more realistic simulations, more accurate distributions of Einstein radii and faint lens galaxies, will give a more accurate account of how CNNs will perform with real data. Machine learning techniques will provide a subset of ostensibly reliable lens systems where verification by visual inspection can be achieved in a realistic time$-$scale which can then be used to refine the CNN.

\section*{Acknowledgements}

We thank the anonymous referee for many helpful and constructive comments. We thank the Science and Technology Facilities Council for financial support under grants ST/N50421X/1 and ST/P000584/1. We acknowledge support during the preparation of this work from the International Space Science Institute (ISSI), Berne, Switzerland, in the form of support for meetings of the collaboration 'Strong Gravitational Lensing with Current and Future Space Observations', P.I. J-P. Kneib.

%%%%%%%%%%%%%%%%%%%%%%%%%%%%%%%%%%%%%%%%%%%%%%%%%%



%\bibliographystyle{mnras}

\bibliographystyle{mnras}
\bibliography{References}

\begin{thebibliography}{}
\makeatletter
\relax
\def\mn@urlcharsother{\let\do\@makeother \do\$\do\&\do\#\do\^\do\_\do\%\do\~}
\def\mn@doi{\begingroup\mn@urlcharsother \@ifnextchar [ {\mn@doi@}
  {\mn@doi@[]}}
\def\mn@doi@[#1]#2{\def\@tempa{#1}\ifx\@tempa\@empty \href
  {http://dx.doi.org/#2} {doi:#2}\else \href {http://dx.doi.org/#2} {#1}\fi
  \endgroup}
\def\mn@eprint#1#2{\mn@eprint@#1:#2::\@nil}
\def\mn@eprint@arXiv#1{\href {http://arxiv.org/abs/#1} {{\tt arXiv:#1}}}
\def\mn@eprint@dblp#1{\href {http://dblp.uni-trier.de/rec/bibtex/#1.xml}
  {dblp:#1}}
\def\mn@eprint@#1:#2:#3:#4\@nil{\def\@tempa {#1}\def\@tempb {#2}\def\@tempc
  {#3}\ifx \@tempc \@empty \let \@tempc \@tempb \let \@tempb \@tempa \fi \ifx
  \@tempb \@empty \def\@tempb {arXiv}\fi \@ifundefined
  {mn@eprint@\@tempb}{\@tempb:\@tempc}{\expandafter \expandafter \csname
  mn@eprint@\@tempb\endcsname \expandafter{\@tempc}}}

\bibitem[\protect\citeauthoryear{Aggarwal}{Aggarwal}{2014}]{Aggarwal2014}
Aggarwal C.~C.,  2014, Data classification: algorithms and applications..
Chapman \& Hall, CRC Press

\bibitem[\protect\citeauthoryear{Amendola et~al.,}{Amendola
  et~al.}{2018}]{Amendola2018}
Amendola L.,  et~al., 2018, \mn@doi [Living Reviews in Relativity]
  {10.1007/s41114-017-0010-3}, \href
  {http://adsabs.harvard.edu/abs/2018LRR....21....2A} {21, 2}

\bibitem[\protect\citeauthoryear{Blain}{Blain}{1996}]{Blain1996}
Blain A.,  1996, \mn@doi [mnras] {10.1093/mnras/283.4.1340}, \href
  {http://adsabs.harvard.edu/abs/1996MNRAS.283.1340B} {283, 1340}

\bibitem[\protect\citeauthoryear{{Bolton}, {Burles}, {Koopmans}, {Treu},
  {Gavazzi}, {Moustakas}, {Wayth}  \& {Schlegel}}{{Bolton}
  et~al.}{2008}]{Bolton2008}
{Bolton} A.~S.,  {Burles} S.,  {Koopmans} L.~V.~E.,  {Treu} T.,  {Gavazzi} R.,
  {Moustakas} L.~A.,  {Wayth} R.,   {Schlegel} D.~J.,  2008, \mn@doi [\apj]
  {10.1086/589327}, \href {http://adsabs.harvard.edu/abs/2008ApJ...682..964B}
  {682, 964}

\bibitem[\protect\citeauthoryear{Browne, Jackson, Augusto, Henstock, Marlow,
  Nair  \& Wilkinson}{Browne et~al.}{1997}]{Browne1997}
Browne I.,  Jackson N.,  Augusto P.,  Henstock D.,  Marlow D.,  Nair S.,
  Wilkinson P.,  1997

\bibitem[\protect\citeauthoryear{Browne et~al.,}{Browne
  et~al.}{2003}]{Browne2003}
Browne I. W.~A.,  et~al., 2003, \mn@doi [Monthly Notices of the Royal
  Astronomical Society] {10.1046/j.1365-8711.2003.06257.x}, 341, 13

\bibitem[\protect\citeauthoryear{{Brownstein} et~al.,}{{Brownstein}
  et~al.}{2012}]{Brownstein2012}
{Brownstein} J.~R.,  et~al., 2012, \mn@doi [\apj] {10.1088/0004-637X/744/1/41},
  \href {http://adsabs.harvard.edu/abs/2012ApJ...744...41B} {744, 41}

\bibitem[\protect\citeauthoryear{Chae}{Chae}{2003}]{Chae2003}
Chae K.-H.,  2003, \mn@doi [Monthly Notices of the Royal Astronomical Society]
  {10.1111/j.1365-2966.2003.07092.x}, 346, 746

\bibitem[\protect\citeauthoryear{Ciresan, Giusti, Gambardella  \&
  Schmidhuber}{Ciresan et~al.}{2013}]{Ciresan2013}
Ciresan D.~C.,  Giusti A.,  Gambardella L.~M.,   Schmidhuber J.,  2013, in
  MICCAI. pp 411--418

\bibitem[\protect\citeauthoryear{Claeskens, Sluse, Riaud  \& Surdej}{Claeskens
  et~al.}{2006}]{Claeskens2006}
Claeskens J.-F.,  Sluse D.,  Riaud P.,   Surdej J.,  2006, \mn@doi [A\&A]
  {10.1051/0004-6361:20054352}, 451, 865

\bibitem[\protect\citeauthoryear{Collett}{Collett}{2015}]{Collett2015}
Collett T.~E.,  2015, The Astrophysical Journal, 811, 20

\bibitem[\protect\citeauthoryear{Collett \& Auger}{Collett \&
  Auger}{2014}]{CollettandAuger2014}
Collett T.~E.,  Auger M.~W.,  2014, \mn@doi [Monthly Notices of the Royal
  Astronomical Society] {10.1093/mnras/stu1190}, 443, 969

\bibitem[\protect\citeauthoryear{{Dom{\'{\i}}nguez S{\'a}nchez}
  et~al.,}{{Dom{\'{\i}}nguez S{\'a}nchez} et~al.}{2018}]{Relearn}
{Dom{\'{\i}}nguez S{\'a}nchez} H.,  et~al., 2018, preprint, \href
  {http://adsabs.harvard.edu/abs/2018arXiv180700807D} {} (\mn@eprint {arXiv}
  {1807.00807})

\bibitem[\protect\citeauthoryear{Faure et~al.,}{Faure et~al.}{2008}]{Faure2008}
Faure C.,  et~al., 2008, The Astrophysical Journal Supplement Series, 178, 382

\bibitem[\protect\citeauthoryear{{Gavazzi}, {Treu}, {Koopmans}, {Bolton},
  {Moustakas}, {Burles}  \& {Marshall}}{{Gavazzi} et~al.}{2008}]{Gavazzi2008}
{Gavazzi} R.,  {Treu} T.,  {Koopmans} L.~V.~E.,  {Bolton} A.~S.,  {Moustakas}
  L.~A.,  {Burles} S.,   {Marshall} P.~J.,  2008, \mn@doi [\apj]
  {10.1086/529541}, \href {http://adsabs.harvard.edu/abs/2008ApJ...677.1046G}
  {677, 1046}

\bibitem[\protect\citeauthoryear{Gavazzi, Marshall, Treu  \&
  Sonnenfeld}{Gavazzi et~al.}{2014}]{Gavazzi2014}
Gavazzi R.,  Marshall P.~J.,  Treu T.,   Sonnenfeld A.,  2014, The
  Astrophysical Journal, 785, 144

\bibitem[\protect\citeauthoryear{He, Zhang, Ren  \& Sun}{He
  et~al.}{2015}]{HeNormal}
He K.,  Zhang X.,  Ren S.,   Sun J.,  2015, in The IEEE International
  Conference on Computer Vision (ICCV).

\bibitem[\protect\citeauthoryear{{Jackson}}{{Jackson}}{2008}]{Jackson2008}
{Jackson} N.,  2008, \mn@doi [\mnras] {10.1111/j.1365-2966.2008.13629.x}, \href
  {http://adsabs.harvard.edu/abs/2008MNRAS.389.1311J} {389, 1311}

\bibitem[\protect\citeauthoryear{{Jackson}}{{Jackson}}{2011}]{Jackson2011}
{Jackson} N.,  2011, \mn@doi [\apjl] {10.1088/2041-8205/739/1/L28}, \href
  {http://adsabs.harvard.edu/abs/2011ApJ...739L..28J} {739, L28}

\bibitem[\protect\citeauthoryear{Kim}{Kim}{2007}]{Kim2016}
Kim J.,  2007, \mn@doi [Monthly Notices of the Royal Astronomical Society]
  {10.1111/j.1365-2966.2006.11285.x}, 375, 625

\bibitem[\protect\citeauthoryear{Kingma \& Ba}{Kingma \& Ba}{2014}]{Kingma2014}
Kingma D.~P.,  Ba J.,  2014, CoRR, abs/1412.6980

\bibitem[\protect\citeauthoryear{{Kochanek} \& {Schechter}}{{Kochanek} \&
  {Schechter}}{2004}]{Kochanek2004}
{Kochanek} C.~S.,  {Schechter} P.~L.,  2004, Measuring and Modeling the
  Universe, \href {http://adsabs.harvard.edu/abs/2004mmu..symp..117K} {p.~117}

\bibitem[\protect\citeauthoryear{{LSST Science Collaboration} et~al.,}{{LSST
  Science Collaboration} et~al.}{2009}]{LSSTScienceCollaboration2009}
{LSST Science Collaboration} et~al., 2009, preprint, \href
  {http://adsabs.harvard.edu/abs/2009arXiv0912.0201L} {} (\mn@eprint {arXiv}
  {0912.0201})

\bibitem[\protect\citeauthoryear{{Laureijs} et~al.,}{{Laureijs}
  et~al.}{2011}]{Laureijs2011}
{Laureijs} R.,  et~al., 2011, preprint, \href
  {http://adsabs.harvard.edu/abs/2011arXiv1110.3193L} {} (\mn@eprint {arXiv}
  {1110.3193})

\bibitem[\protect\citeauthoryear{LeCun, Bengio  \& Hinton}{LeCun
  et~al.}{2015}]{LeCun2015}
LeCun Y.,  Bengio Y.,   Hinton G.,  2015, Nature, 521, 436

\bibitem[\protect\citeauthoryear{{Lenzen, F.}, {Schindler, S.}  \& {Scherzer,
  O.}}{{Lenzen, F.} et~al.}{2004}]{Lenzen2004}
{Lenzen, F.} {Schindler, S.}  {Scherzer, O.} 2004, \mn@doi [A\&A]
  {10.1051/0004-6361:20034619}, 416, 391

\bibitem[\protect\citeauthoryear{{Mallat}}{{Mallat}}{2016}]{Mallat2016}
{Mallat} S.,  2016, \mn@doi [Philosophical Transactions of the Royal Society of
  London Series A] {10.1098/rsta.2015.0203}, \href
  {http://adsabs.harvard.edu/abs/2016RSPTA.37450203M} {374, 20150203}

\bibitem[\protect\citeauthoryear{{Marchetti}, {Serjeant}  \&
  {Vaccari}}{{Marchetti} et~al.}{2017}]{Marchetti2017}
{Marchetti} L.,  {Serjeant} S.,   {Vaccari} M.,  2017, \mn@doi [\mnras]
  {10.1093/mnras/stx1553}, \href
  {http://adsabs.harvard.edu/abs/2017MNRAS.470.5007M} {470, 5007}

\bibitem[\protect\citeauthoryear{{Marshall} et~al.,}{{Marshall}
  et~al.}{2016}]{Marshall2016}
{Marshall} P.~J.,  et~al., 2016, \mn@doi [\mnras] {10.1093/mnras/stv2009},
  \href {http://adsabs.harvard.edu/abs/2016MNRAS.455.1171M} {455, 1171}

\bibitem[\protect\citeauthoryear{{McKean} et~al.,}{{McKean}
  et~al.}{2015}]{McKean2015}
{McKean} J.,  et~al., 2015, Advancing Astrophysics with the Square Kilometre
  Array (AASKA14), \href {http://adsabs.harvard.edu/abs/2015aska.confE..84M}
  {p.~84}

\bibitem[\protect\citeauthoryear{{Metcalf} \& {Madau}}{{Metcalf} \&
  {Madau}}{2001}]{Metcalf2001}
{Metcalf} R.~B.,  {Madau} P.,  2001, \mn@doi [\apj] {10.1086/323695}, \href
  {http://adsabs.harvard.edu/abs/2001ApJ...563....9M} {563, 9}

\bibitem[\protect\citeauthoryear{Metcalf \& Petkova}{Metcalf \&
  Petkova}{2014}]{Metcalf2014}
Metcalf R.~B.,  Petkova M.,  2014, \mn@doi [Monthly Notices of the Royal
  Astronomical Society] {10.1093/mnras/stu1859}, 445, 1942

\bibitem[\protect\citeauthoryear{{Metcalf} et~al.,}{{Metcalf}
  et~al.}{2018}]{LensChallenge}
{Metcalf} R.~B.,  et~al., 2018, preprint, \href
  {http://adsabs.harvard.edu/abs/2018arXiv180203609M} {} (\mn@eprint {arXiv}
  {1802.03609})

\bibitem[\protect\citeauthoryear{Miyazaki et~al.,}{Miyazaki
  et~al.}{2012}]{Miyazaki2012}
Miyazaki S.,  et~al., 2012, Hyper Suprime-Cam, \mn@doi{10.1117/12.926844}, \url
  {https://doi.org/10.1117/12.926844}

\bibitem[\protect\citeauthoryear{More, Cabanac, More, Alard, Limousin, Kneib,
  Gavazzi  \& Motta}{More et~al.}{2012}]{More2012}
More A.,  Cabanac R.,  More S.,  Alard C.,  Limousin M.,  Kneib J.-P.,  Gavazzi
  R.,   Motta V.,  2012, The Astrophysical Journal, 749, 38

\bibitem[\protect\citeauthoryear{More et~al.,}{More et~al.}{2016}]{More2016}
More A.,  et~al., 2016, \mn@doi [Monthly Notices of the Royal Astronomical
  Society] {10.1093/mnras/stv1965}, 455, 1191

\bibitem[\protect\citeauthoryear{{Myers} et~al.,}{{Myers}
  et~al.}{2003}]{Myers2003}
{Myers} S.~T.,  et~al., 2003, \mn@doi [\mnras]
  {10.1046/j.1365-8711.2003.06256.x}, \href
  {http://adsabs.harvard.edu/abs/2003MNRAS.341....1M} {341, 1}

\bibitem[\protect\citeauthoryear{{Oguri} \& {Marshall}}{{Oguri} \&
  {Marshall}}{2010}]{Oguri2010}
{Oguri} M.,  {Marshall} P.~J.,  2010, \mn@doi [\mnras]
  {10.1111/j.1365-2966.2010.16639.x}, \href
  {http://adsabs.harvard.edu/abs/2010MNRAS.405.2579O} {405, 2579}

\bibitem[\protect\citeauthoryear{{Patnaik}, {Browne}, {Walsh}, {Chaffee}  \&
  {Foltz}}{{Patnaik} et~al.}{1992}]{Patnaik1992}
{Patnaik} A.~R.,  {Browne} I.~W.~A.,  {Walsh} D.,  {Chaffee} F.~H.,   {Foltz}
  C.~B.,  1992, \mn@doi [\mnras] {10.1093/mnras/259.1.1P}, \href
  {http://adsabs.harvard.edu/abs/1992MNRAS.259P...1P} {259, 1P}

\bibitem[\protect\citeauthoryear{{Perley} et~al.,}{{Perley}
  et~al.}{2009}]{Perley2009}
{Perley} R.,  et~al., 2009, \mn@doi [IEEE Proceedings]
  {10.1109/JPROC.2009.2015470}, \href
  {http://adsabs.harvard.edu/abs/2009IEEEP..97.1448P} {97, 1448}

\bibitem[\protect\citeauthoryear{{Petrillo} et~al.,}{{Petrillo}
  et~al.}{2017}]{Petrillo2017}
{Petrillo} C.~E.,  et~al., 2017, \mn@doi [\mnras] {10.1093/mnras/stx2052},
  \href {http://adsabs.harvard.edu/abs/2017MNRAS.472.1129P} {472, 1129}

\bibitem[\protect\citeauthoryear{{Rawlings} \& {Schilizzi}}{{Rawlings} \&
  {Schilizzi}}{2011}]{Rawlings2011}
{Rawlings} S.,  {Schilizzi} R.,  2011, preprint, \href
  {http://adsabs.harvard.edu/abs/2011arXiv1105.5953R} {} (\mn@eprint {arXiv}
  {1105.5953})

\bibitem[\protect\citeauthoryear{{Refsdal}}{{Refsdal}}{1964}]{Refsdal1964}
{Refsdal} S.,  1964, \mn@doi [\mnras] {10.1093/mnras/128.4.295}, \href
  {http://adsabs.harvard.edu/abs/1964MNRAS.128..295R} {128, 295}

\bibitem[\protect\citeauthoryear{{Seidel, G.} \& {Bartelmann, M.}}{{Seidel, G.}
  \& {Bartelmann, M.}}{2007}]{Seidel2007}
{Seidel, G.} {Bartelmann, M.} 2007, \mn@doi [A\&A]
  {10.1051/0004-6361:20066097}, 472, 341

\bibitem[\protect\citeauthoryear{{Serjeant}}{{Serjeant}}{2014}]{Serjeant2014}
{Serjeant} S.,  2014, \mn@doi [\apjl] {10.1088/2041-8205/793/1/L10}, \href
  {http://adsabs.harvard.edu/abs/2014ApJ...793L..10S} {793, L10}

\bibitem[\protect\citeauthoryear{{The Dark Energy Survey Collaboration}}{{The
  Dark Energy Survey Collaboration}}{2005}]{DarkEnergySurveyCollaboration2005}
{The Dark Energy Survey Collaboration} 2005, ArXiv Astrophysics e-prints, \href
  {http://adsabs.harvard.edu/abs/2005astro.ph.10346T} {}

\bibitem[\protect\citeauthoryear{Thorpe, Fize  \& Marlot}{Thorpe
  et~al.}{1996}]{Thorpe1996}
Thorpe S.,  Fize D.,   Marlot C.,  1996, Nature, 381, 520

\bibitem[\protect\citeauthoryear{Treu \& Koopmans}{Treu \&
  Koopmans}{2002}]{Treu2002}
Treu T.,  Koopmans L. V.~E.,  2002, \mn@doi [Monthly Notices of the Royal
  Astronomical Society] {10.1046/j.1365-8711.2002.06107.x}, 337, L6

\bibitem[\protect\citeauthoryear{{Vegetti}, {Lagattuta}, {McKean}, {Auger},
  {Fassnacht}  \& {Koopmans}}{{Vegetti} et~al.}{2012}]{Vegetti2012}
{Vegetti} S.,  {Lagattuta} D.~J.,  {McKean} J.~P.,  {Auger} M.~W.,  {Fassnacht}
  C.~D.,   {Koopmans} L.~V.~E.,  2012, \mn@doi [\nat] {10.1038/nature10669},
  \href {http://adsabs.harvard.edu/abs/2012Natur.481..341V} {481, 341}

\bibitem[\protect\citeauthoryear{{Walsh}, {Carswell}  \& {Weymann}}{{Walsh}
  et~al.}{1979}]{Walsh1979}
{Walsh} D.,  {Carswell} R.~F.,   {Weymann} R.~J.,  1979, \mn@doi [\nat]
  {10.1038/279381a0}, \href {http://adsabs.harvard.edu/abs/1979Natur.279..381W}
  {279, 381}

\bibitem[\protect\citeauthoryear{{Wyithe}, {Yan}, {Windhorst}  \&
  {Mao}}{{Wyithe} et~al.}{2011}]{Wyithe2011}
{Wyithe} J.~S.~B.,  {Yan} H.,  {Windhorst} R.~A.,   {Mao} S.,  2011, \mn@doi
  [\nat] {10.1038/nature09619}, \href
  {http://adsabs.harvard.edu/abs/2011Natur.469..181W} {469, 181}

\bibitem[\protect\citeauthoryear{{de Jong, Jelte T. A.} et~al.,}{{de Jong,
  Jelte T. A.} et~al.}{2015}]{deJong2015}
{de Jong, Jelte T. A.} et~al., 2015, \mn@doi [A\&A]
  {10.1051/0004-6361/201526601}, 582, A62

\makeatother
\end{thebibliography}

%%%%%%%%%%%%%%%%%%%%%%%%%%%%%%%%%%%%%%%%%%%%%%%%%%

%%%%%%%%%%%%%%%%% APPENDICES %%%%%%%%%%%%%%%%%%%%%

\appendix

\section{COSMOS Lenses}
The following figures show the 65 lenses from COSMOS that our CNNs were tested with. Each lens has the ID number below. Images with an asterisk after the ID number were the ones that were correctly identified as a lens. Each image has the usual North up, East left configuration, and are 10$ \times $10 arcseconds. Images in the left column are from the COSMOS survey, in the right are the same images after being convolved with a kernel to give the image the same PSF as Euclid. The images have been grayscale altered to best show the lens.

\begin{figure}
\includegraphics[width=.8in]{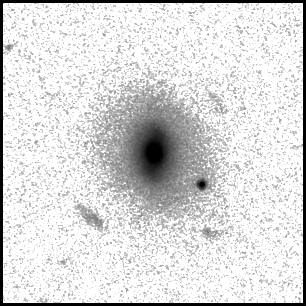}
\includegraphics[width=.8in]{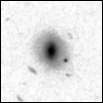}
\includegraphics[width=.8in]{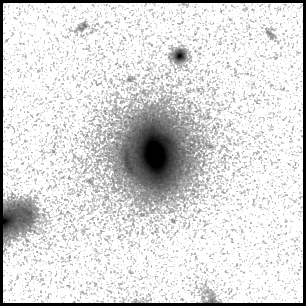}
\includegraphics[width=.8in]{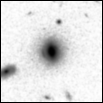}
\centering
0009+2455 \tab \tab \tab 0012+2015
\end{figure}
\begin{figure}
\includegraphics[width=.8in]{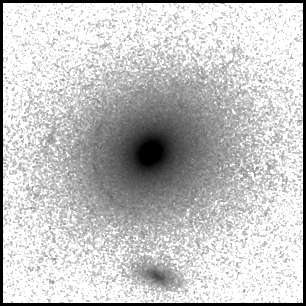}
\includegraphics[width=.8in]{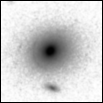}
\includegraphics[width=.8in]{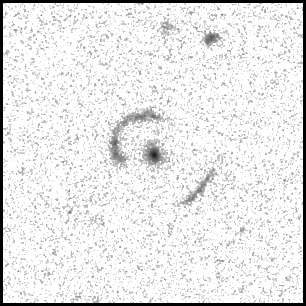}
\includegraphics[width=.8in]{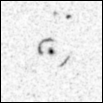}
\centering
0013+2249 \tab \tab \tab 0018+3845*
\end{figure}
\begin{figure}
\includegraphics[width=.8in]{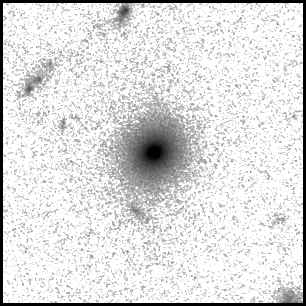}
\includegraphics[width=.8in]{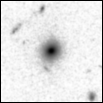}
\includegraphics[width=.8in]{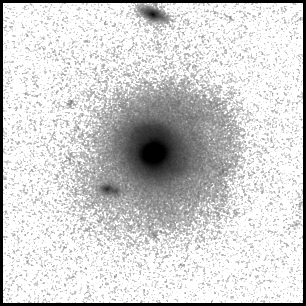}
\includegraphics[width=.8in]{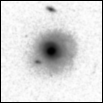}
\centering
0028+1919 \tab \tab \tab 0029+4018
\end{figure}
\begin{figure}
\includegraphics[width=.8in]{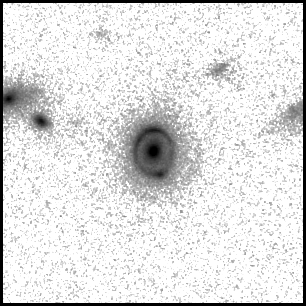}
\includegraphics[width=.8in]{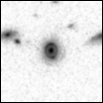}
\includegraphics[width=.8in]{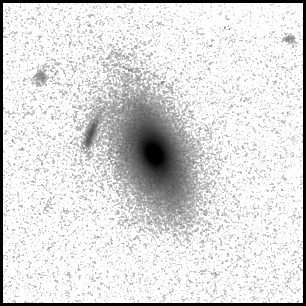}
\includegraphics[width=.8in]{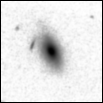}
\centering
0038+4133* \tab \tab \tab 0047+2931
\end{figure}
\begin{figure}
\includegraphics[width=.8in]{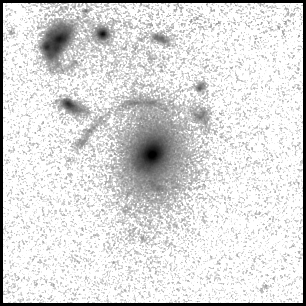}
\includegraphics[width=.8in]{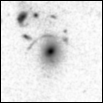}
\includegraphics[width=.8in]{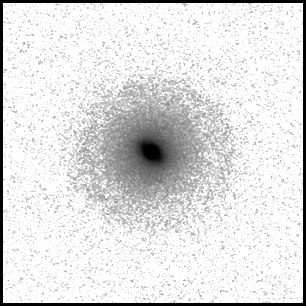}
\includegraphics[width=.8in]{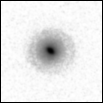}
\centering
0047+5023 \tab \tab \tab 0049+5128*
\end{figure}
\begin{figure}
\includegraphics[width=.8in]{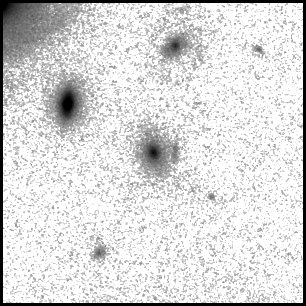}
\includegraphics[width=.8in]{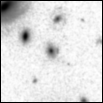}
\includegraphics[width=.8in]{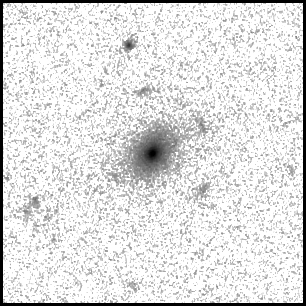}
\includegraphics[width=.8in]{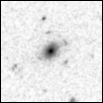}
\centering
0050+0357* \tab \tab \tab 0050+4901
\end{figure}
\begin{figure}
\includegraphics[width=.8in]{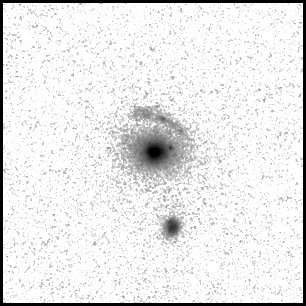}
\includegraphics[width=.8in]{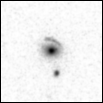}
\includegraphics[width=.8in]{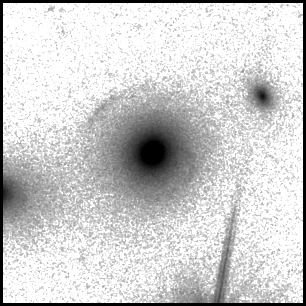}
\includegraphics[width=.8in]{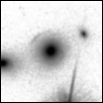}
\centering
0055+3821* \tab \tab \tab 0056+1226*
\end{figure}
\begin{figure}
\includegraphics[width=.8in]{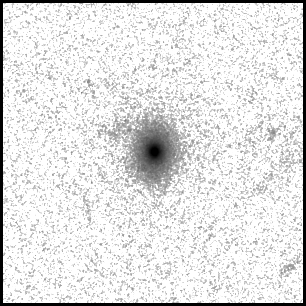}
\includegraphics[width=.8in]{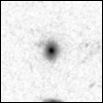}
\includegraphics[width=.8in]{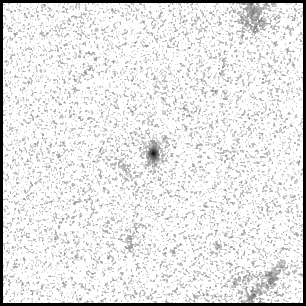}
\includegraphics[width=.8in]{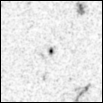}
\centering
0056+2106 \tab \tab \tab 0104+2046
\end{figure}
\begin{figure}
\includegraphics[width=.8in]{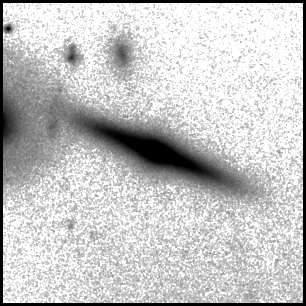}
\includegraphics[width=.8in]{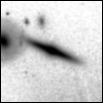}
\includegraphics[width=.8in]{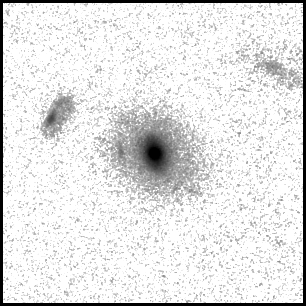}
\includegraphics[width=.8in]{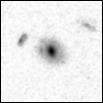}
\centering
0104+2501* \tab \tab \tab 0105+4531*
\end{figure}
\begin{figure}
\includegraphics[width=.8in]{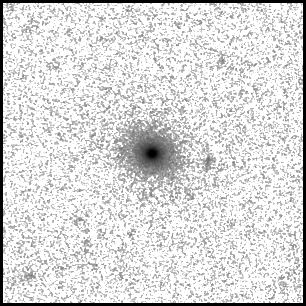}
\includegraphics[width=.8in]{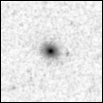}
\includegraphics[width=.8in]{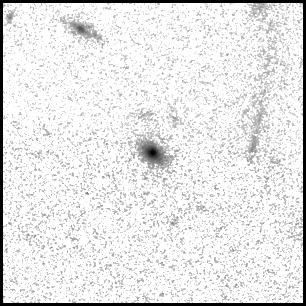}
\includegraphics[width=.8in]{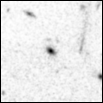}
\centering
0107+0533 \tab \tab \tab 0108+5606
\end{figure}
\begin{figure}
\includegraphics[width=.8in]{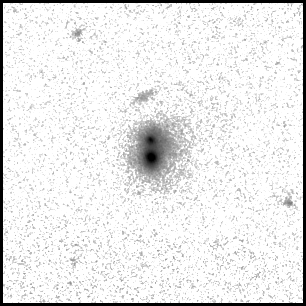}
\includegraphics[width=.8in]{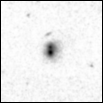}
\includegraphics[width=.8in]{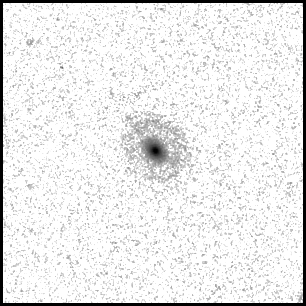}
\includegraphics[width=.8in]{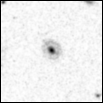}
\centering
0120+4551 \tab \tab \tab 0124+5121*
\end{figure}
\begin{figure}
\includegraphics[width=.8in]{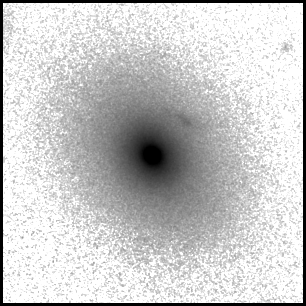}
\includegraphics[width=.8in]{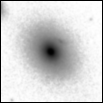}
\includegraphics[width=.8in]{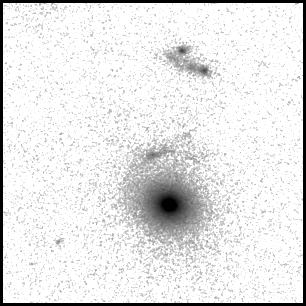}
\includegraphics[width=.8in]{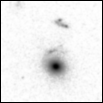}
\centering
0148+2325 \tab \tab \tab 0208+1422*
\end{figure}
\begin{figure}
\includegraphics[width=.8in]{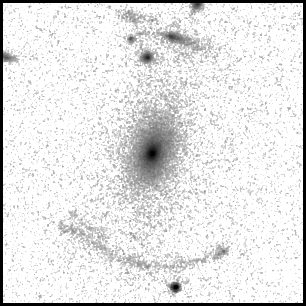}
\includegraphics[width=.8in]{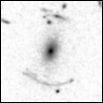}
\includegraphics[width=.8in]{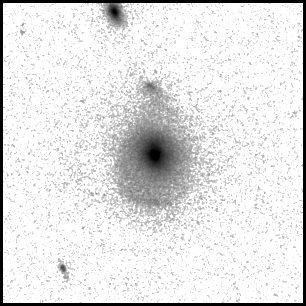}
\includegraphics[width=.8in]{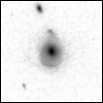}
\centering
0211+1139* \tab \tab \tab 0216+2955
\end{figure}
\begin{figure}
\includegraphics[width=.8in]{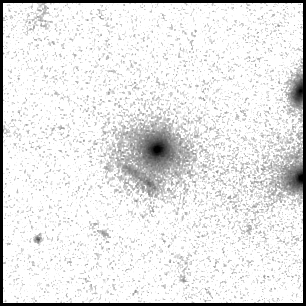}
\includegraphics[width=.8in]{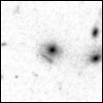}
\includegraphics[width=.8in]{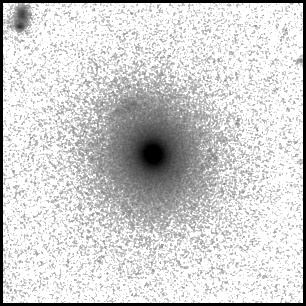}
\includegraphics[width=.8in]{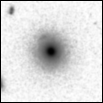}
\centering
0220+2335* \tab \tab \tab 0221+3440
\end{figure}
\begin{figure}
\includegraphics[width=.8in]{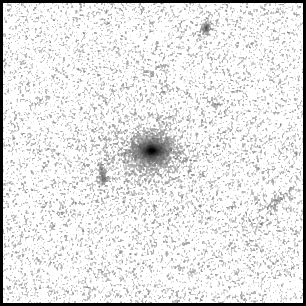}
\includegraphics[width=.8in]{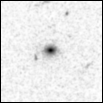}
\includegraphics[width=.8in]{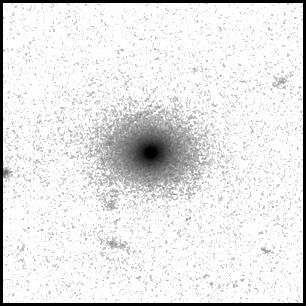}
\includegraphics[width=.8in]{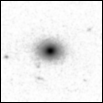}
\centering
0227+0451* \tab \tab \tab 0236+4807
\end{figure}
\begin{figure}
\includegraphics[width=.8in]{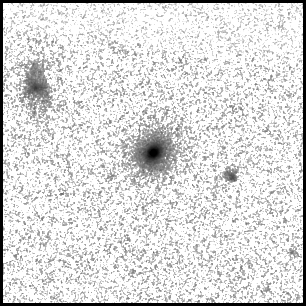}
\includegraphics[width=.8in]{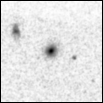}
\includegraphics[width=.8in]{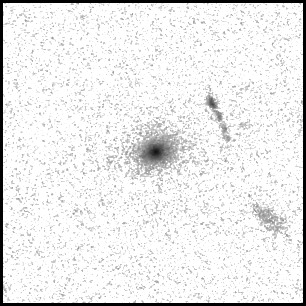}
\includegraphics[width=.8in]{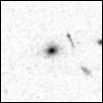}
\centering
0248+1422 \tab \tab \tab 5748+5524
\end{figure}
\begin{figure}
\includegraphics[width=.8in]{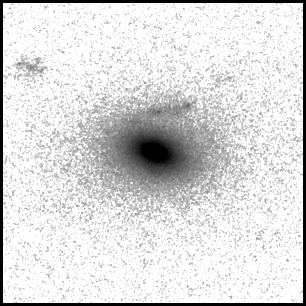}
\includegraphics[width=.8in]{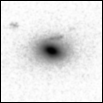}
\includegraphics[width=.8in]{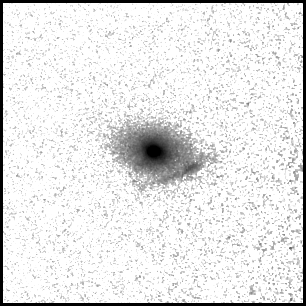}
\includegraphics[width=.8in]{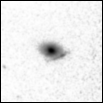}
\centering
5750+5619 \tab \tab \tab 5752+2057
\end{figure}
\begin{figure}
\includegraphics[width=.8in]{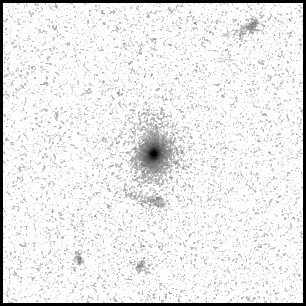}
\includegraphics[width=.8in]{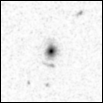}
\includegraphics[width=.8in]{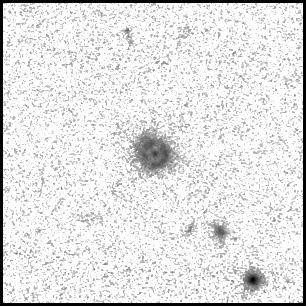}
\includegraphics[width=.8in]{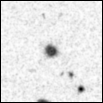}
\centering
5754+5952 \tab \tab \tab 5758+1525*
\end{figure}
\begin{figure}
\includegraphics[width=.8in]{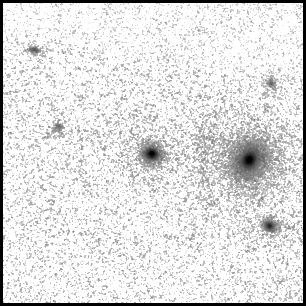}
\includegraphics[width=.8in]{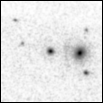}
\includegraphics[width=.8in]{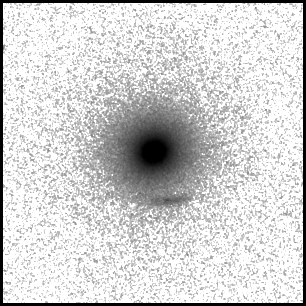}
\includegraphics[width=.8in]{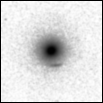}
\centering
5805+0413* \tab \tab \tab 5806+5809
\end{figure}
\begin{figure}
\includegraphics[width=.8in]{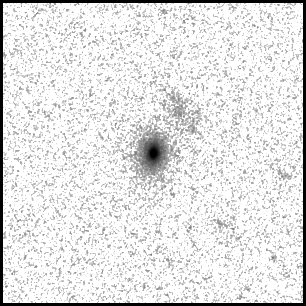}
\includegraphics[width=.8in]{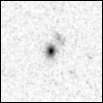}
\includegraphics[width=.8in]{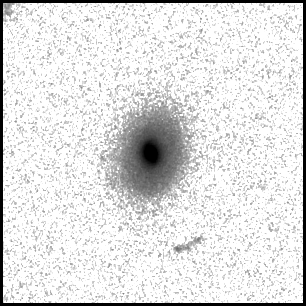}
\includegraphics[width=.8in]{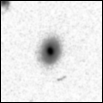}
\centering
5821+4437 \tab \tab \tab 5829+3734
\end{figure}
\begin{figure}
\includegraphics[width=.8in]{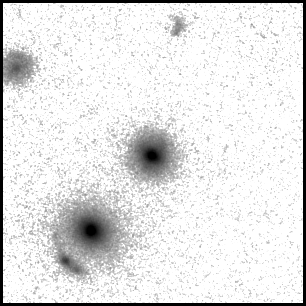}
\includegraphics[width=.8in]{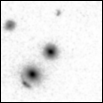}
\includegraphics[width=.8in]{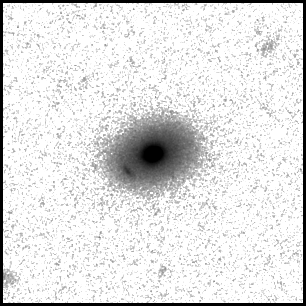}
\includegraphics[width=.8in]{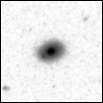}
\centering
5831+4332* \tab \tab \tab 5841+4646
\end{figure}
\begin{figure}
\includegraphics[width=.8in]{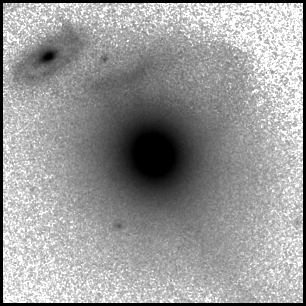}
\includegraphics[width=.8in]{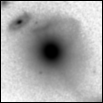}
\includegraphics[width=.8in]{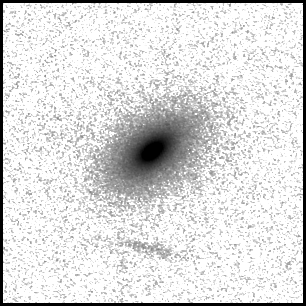}
\includegraphics[width=.8in]{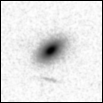}
\centering
5851+1813 \tab \tab \tab 5856+4755
\end{figure}
\begin{figure}
\includegraphics[width=.8in]{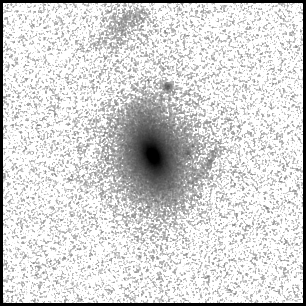}
\includegraphics[width=.8in]{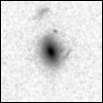}
\includegraphics[width=.8in]{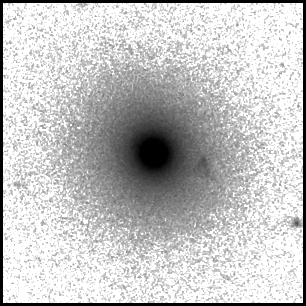}
\includegraphics[width=.8in]{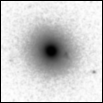}
\centering
5857+5949 \tab \tab \tab 5906+4524
\end{figure}
\begin{figure}
\includegraphics[width=.8in]{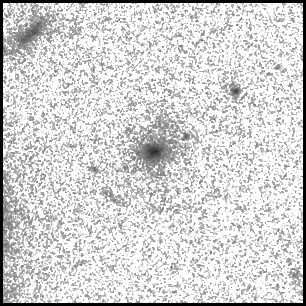}
\includegraphics[width=.8in]{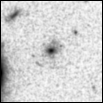}
\includegraphics[width=.8in]{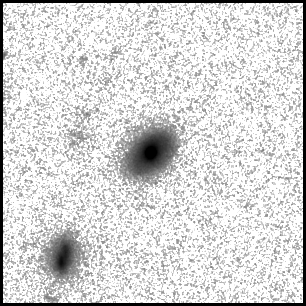}
\includegraphics[width=.8in]{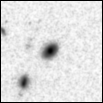}
\centering
5914+1219* \tab \tab \tab 5919+3853
\end{figure}
\begin{figure}
\includegraphics[width=.8in]{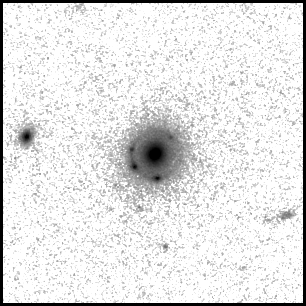}
\includegraphics[width=.8in]{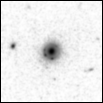}
\includegraphics[width=.8in]{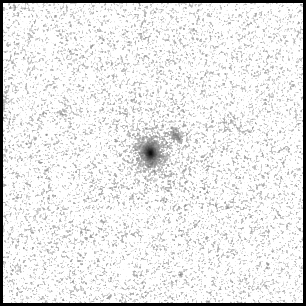}
\includegraphics[width=.8in]{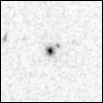}
\centering
5921+0638* \tab \tab \tab 5924+0852
\end{figure}
\begin{figure}
\includegraphics[width=.8in]{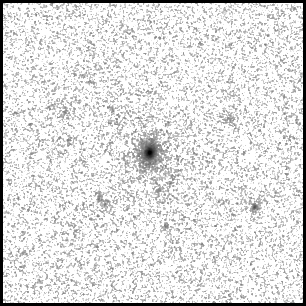}
\includegraphics[width=.8in]{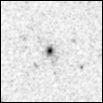}
\includegraphics[width=.8in]{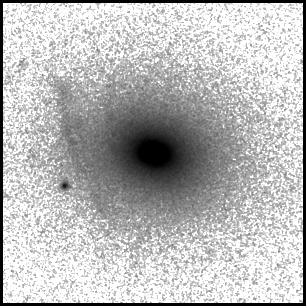}
\includegraphics[width=.8in]{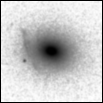}
\centering
5925+3039 \tab \tab \tab 5929+1352
\end{figure}
\begin{figure}
\includegraphics[width=.8in]{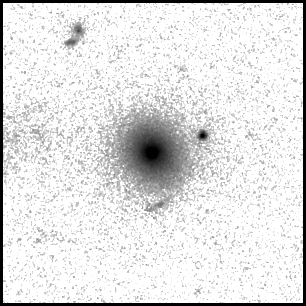}
\includegraphics[width=.8in]{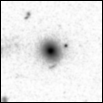}
\includegraphics[width=.8in]{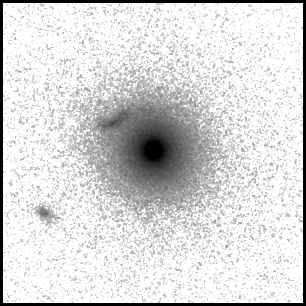}
\includegraphics[width=.8in]{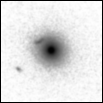}
\centering
5929+4553* \tab \tab \tab 5931+0229
\end{figure}
\begin{figure}
\includegraphics[width=.8in]{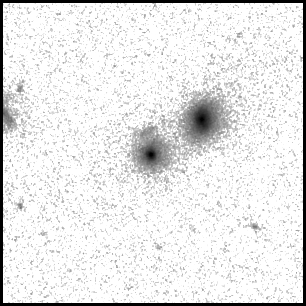}
\includegraphics[width=.8in]{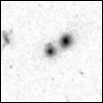}
\includegraphics[width=.8in]{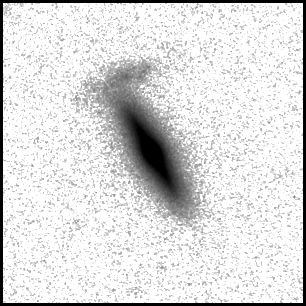}
\includegraphics[width=.8in]{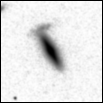}
\centering
5932+1018* \tab \tab \tab 5936+3621
\end{figure}
\begin{figure}
\includegraphics[width=.8in]{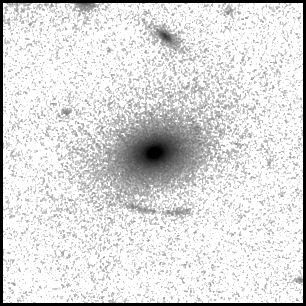}
\includegraphics[width=.8in]{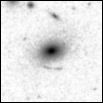}
\includegraphics[width=.8in]{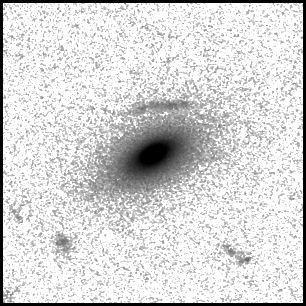}
\includegraphics[width=.8in]{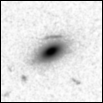}
\centering
5939+3044 \tab \tab \tab 5940+3253
\end{figure}
\begin{figure}
\includegraphics[width=.8in]{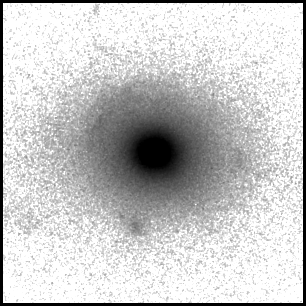}
\includegraphics[width=.8in]{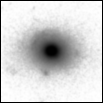}
\includegraphics[width=.8in]{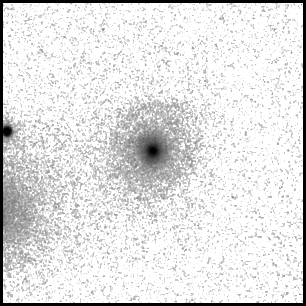}
\includegraphics[width=.8in]{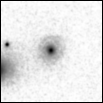}
\centering
5940+5012 \tab \tab \tab 5941+3628*
\end{figure}
\begin{figure}
\includegraphics[width=.8in]{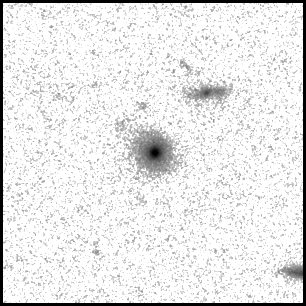}
\includegraphics[width=.8in]{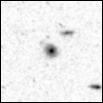}
\includegraphics[width=.8in]{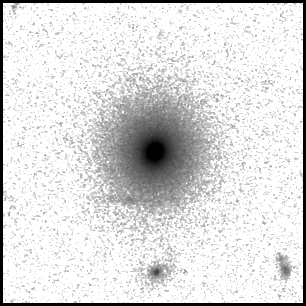}
\includegraphics[width=.8in]{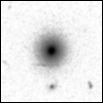}
\centering
5942+2829* \tab \tab \tab 5943+2816
\end{figure}
\begin{figure}
\includegraphics[width=.8in]{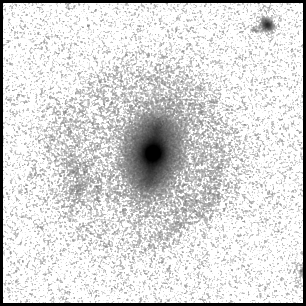}
\includegraphics[width=.8in]{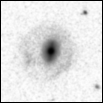}
\includegraphics[width=.8in]{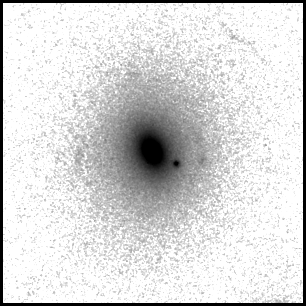}
\includegraphics[width=.8in]{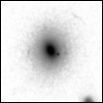}
\centering
5947+4752* \tab \tab \tab 5951+1236
\end{figure}
\begin{figure}
\includegraphics[width=.8in]{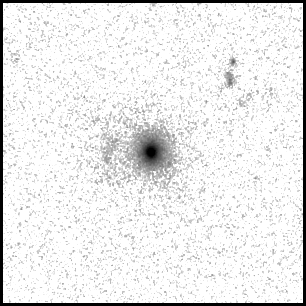}
\includegraphics[width=.8in]{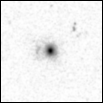}
\includegraphics[width=.8in]{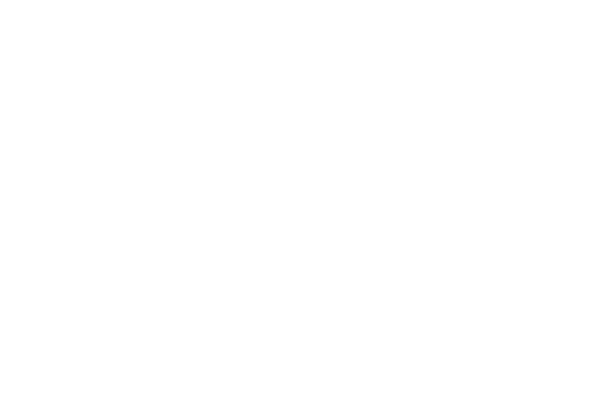}
\includegraphics[width=.8in]{White.png}
\centering
5959+0348 \tab \tab \tab \tab
\end{figure}

%%%%%%%%%%%%%%%%%%%%%%%%%%%%%%%%%%%%%%%%%%%%%%%%%%

% Don't change these lines
\bsp	% typesetting comment
\label{lastpage}
\end{document}